%% file: main.tex
\documentclass[lettersize,journal]{IEEEtran}
\usepackage{amsmath,amsfonts, enumitem}
\usepackage{algorithmic}
\usepackage{algorithm}
\usepackage{array}
\usepackage{textcomp}
\usepackage{stfloats}
\usepackage{url}
\usepackage{verbatim}
\usepackage{cite}
\usepackage[pdftex]{graphicx}
\usepackage{algorithm}
\usepackage{algorithmic}
\usepackage{arydshln}
\usepackage[labelfont=bf, font=small, margin=0pt]{caption}
\usepackage{subcaption}

\usepackage{xcolor}

\begin{document}

\bstctlcite{BSTcontrol}

\title{Domain Expansion via Network Adaptation for Solving Inverse Problems}

\author{Nebiyou Yismaw,
\IEEEmembership{Student Member, IEEE}, 
Ulugbek S. Kamilov,
\IEEEmembership{Senior Member, IEEE}, 
M. Salman Asif,
\IEEEmembership{Senior Member, IEEE} \thanks{This paper is partially based on work supported by the NSF CAREER awards under grants CCF-2043134 and CCF-2046293. }
\thanks{Nebiyou Yismaw and M. Salman Asif are with the University of California Riverside (e-mails: {nyism001@ucr.edu, sasif@ucr.edu}). \\
Ulugbek Kamilov is with Washington University in St. Louis (e-mail: {kamilov@wustl.edu}). 
}
}

\markboth{Domain Expansion via Network Adaptation}{Yismaw \MakeLowercase{\textit{et al.}}}

\maketitle

\begin{abstract}
Deep learning-based methods deliver state-of-the-art performance for solving inverse problems that arise in computational imaging. These methods can be broadly divided into two groups: (1) learn a network to map  measurements to the signal estimate, which is known to be fragile; (2) learn a prior for the signal to use in an optimization-based recovery. Despite the impressive results from the latter approach, many of these methods also lack robustness to shifts in data distribution, measurements, and noise levels. Such domain shifts result in a performance gap and in some cases introduce undesired artifacts in the estimated signal. In this paper, we explore the qualitative and quantitative effects of various domain shifts and propose a flexible and parameter efficient framework that  adapt pretrained networks to such shifts. We demonstrate the effectiveness of our method for a number of natural image, MRI, and CT reconstructions tasks under domain, measurement model, and noise-level shifts. Our experiments demonstrate that our method provides significantly better performance and parameter efficiency compared to existing domain adaptation techniques. 

\end{abstract}

\begin{IEEEkeywords}
Inverse problems, image recovery, domain adaptation, unrolled networks.
\end{IEEEkeywords}

\section{Introduction}

Linear inverse problems arise in many real-world applications. For instance, image enhancement and restoration tasks in denoising, deblurring, and super-resolution or 
medical image reconstruction from indirect measurements in computed tomography (CT) and magnetic resonance imaging (MRI). 
We can model such inverse problems as the recovery of an unknown signal $\mathbf{x}$ from a set of measurements: 
\begin{equation}
    \mathbf{y} = \mathbf{A}\mathbf{x} + \eta, 
    \label{eq:fwd_model}
\end{equation}
where $\mathbf{y}$ represents measurements,  $\mathbf{A}$ represents an $m\times n$ measurement matrix or forward operator, and $\eta$ represents noise. The unknown signal and measurements can be real- or complex-valued. To recover $\mathbf{x}$, we can solve an optimization problem of the following form: 
\begin{equation}
    \label{eq:opti_prob}
\min_{\mathbf{x}} g(\mathbf{x}) + {h}_\theta(\mathbf{x}), 
\end{equation}
where $g(\mathbf{x})$ is a data fidelity term (e.g., $g(\mathbf{x}) = \frac{1}{2} \| \mathbf{y} - \mathbf{A}\mathbf{x} \|^2_2$), $h_\theta(\cdot)$ denotes a regularization function that enforces some prior constraint on the unknown signal, and $\theta$ denotes the regularization function parameters \cite{venkatakrishnan2013plug, boyd2011distributed}. For instance, signal is sparse or low-rank in some representation space or belongs to a manifold of natural images \cite{candes2006stable,mairal2009online, lustig2007sparse, elad2006image}.

\begin{figure*}[t]
    \centering
    \includegraphics[width=\textwidth]{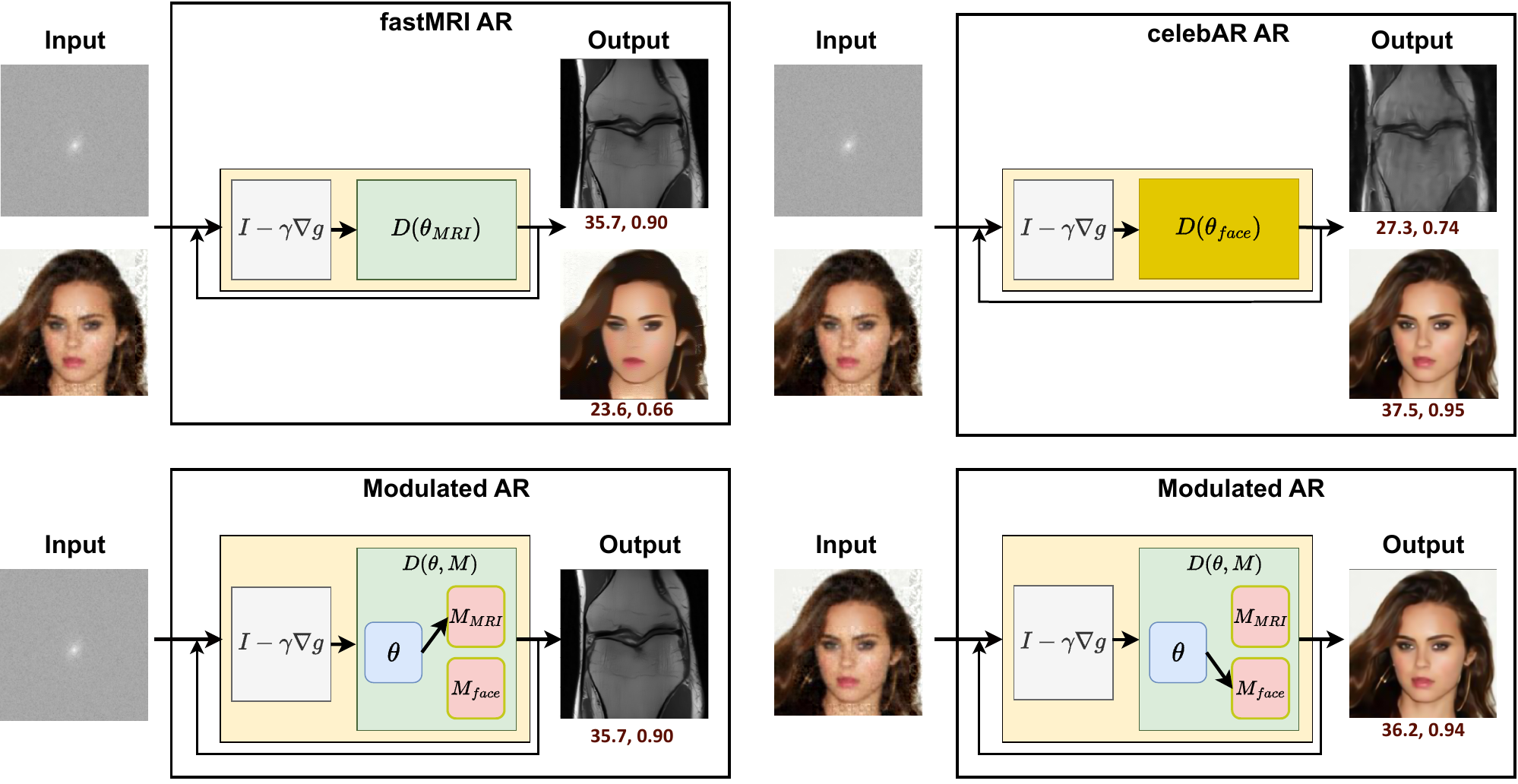}
    \caption{Artifact removal (AR) networks trained on MRI scans (fastMRI AR) and face images (celebA AR) suffer from performance degradation under domain shifts, resulting in poor reconstruction quality (as indicated by PSNR and SSIM values under each image). Our proposed network (Modulated AR) adapts fastMRI AR for face image reconstruction by learning rank-one factors (modulations). The network stores shared and domain-specific modulations separately. During inference, it applies the correct modulation according to the specified domain. Our proposed network retains the performance of fastMRI AR on MR images and achieves competitive reconstruction quality with celebA AR on face images.}
    \label{fig:intro_du_perf_drop_figure}
\end{figure*}

In the deep learning era, we can recover $\mathbf{x}$ by either training a deep (reconstruction) network that maps measurements to the signal estimate or solving an iterative optimization problem (similar to the one in \eqref{eq:opti_prob}) that can also be represented as an unrolled network \cite{ongie2020deep}. 
While training a reconstruction network in an end-to-end manner is possible, it usually requires a large set of input-output training pairs $(\mathbf{y,x})$. Furthermore, since these networks do not explicitly use the forward model in \eqref{eq:fwd_model}, they are known to be sensitive to small changes in the data distribution, measurement operators, and noise \cite{antun2020instabilities, gottschling2020troublesome}. 
Solving the optimization problem in \eqref{eq:opti_prob} with an appropriate choice of regularization function $h(\cdot)$ is often considered a flexible and relatively robust option. 

In recent years, deep networks are often used to represent $h(\cdot)$ instead of hand-designed functions (e.g., $\ell_1$ norm or total variation). For instance, deep unrolling \cite{gregor2010learning, monga2021algorithm,liu2021recovery} and plug-and-play (PnP) \cite{venkatakrishnan2013plug, sun2019online} methods use artifact-removal or image denoising networks that are trained to map a noisy or corrupted estimate of an image onto a clean image manifold \cite{ongie2020deep, monga2021algorithm,liu2021recovery}. 
Despite recent success of such deep unrolled or PnP methods, they are also sensitive to shifts in the data distribution \cite{darestani2021measuring}. 
Figure \ref{fig:intro_du_perf_drop_figure} illustrates this effect for deep unrolling with artifact removal (AR) networks under domain and forward model shifts. The fastMRI AR is trained while solving \eqref{eq:opti_prob} for MR image reconstruction from radially under-sampled k-space measurements. CelebA AR is trained while solving \eqref{eq:opti_prob} to reconstruct face images from measurements obtained using a Gaussian sampling matrix. Note that reconstructing MR images  using the CelebA AR and vice versa results in a significant performance degradation. 

In this paper, we propose a parameter-efficient method to adapt pretrained networks to multiple domains, measurement models, and noise with little to no drop in performance. In particular, we propose a domain-specific modulation of network weights using low-rank (or rank-one) factors. 
Given a single deep unrolled network, we learn a compact set of modulation parameters for each domain/measurement/noise setting, and adapt the weights of the network according to the specific problem at the inference time. In the remainder of the paper, we use the term domain shift and domain adaptation to refer to changes in data/measurement/noise distributions. We present a set of experiments to demonstrate the effectiveness of our method in adapting the deep unrolled network for shifts in data distribution/domain ($\mathbf{x}$), forward models ($\mathbf{A}$), and noise levels ($\eta$).  
The modulated AR in Figure~\ref{fig:intro_du_perf_drop_figure} shows an application of our method, where we adapt a pretrained fastMRI AR to celebA. It applies the learned modulations when recovering celebA images and will use the pretrained weights when reconstructing MRI scans. This network recovers images that qualitatively and quantitatively resemble results of the networks trained for the correct domains. The number of parameters needed to adapt the pretrained network is less than $0.5\%$ of the parameters in the pretrained network.

Our method can be viewed as an example of domain adaptation or domain expansion technique, where we update a network trained for a source domain to perform well on several target domains. Fine-tuning pre-trained networks is a widely used method for domain adaptation but suffers from catastrophic forgetting \cite{mccloskey1989catastrophic} and requires a large number of parameters for every new domain \cite{mallya2018piggyback}. Several 
parameter efficient domain adaptation techniques have been proposed in \cite{alanov2022hyperdomainnet, kanakis2020reparameterizing}. Our method resembles some of these methods in spirit and separates the network into shared and domain-specific modules. By limiting the number of parameters for the domain-specific modules, our method provides a parameter-efficient method to learn multiple tasks and domains. Furthermore, conditional computation is efficient during training and inference compared to independent networks \cite{riquelme2021scaling}.

\noindent \textbf{Contributions. } We summarize the contributions of this paper as follows. 
\begin{itemize}[leftmargin=*,noitemsep,topsep=0pt]
    \item We proposed a simple parameter-efficient domain expansion technique to modulate weights of a pretrained network with rank-one factors. Our method expands the domain of the networks and adapts to a variety of data/model shifts that arise in inverse problems. 
    \item Our method requires a small number of domain-specific parameters (less than $0.5\%$ of a single network) that can be stored separately from the shared network weights. This enables the network to continuously adapt to new domains without forgetting previous knowledge; therefore, we call it domain expansion.  
    \item We present a detailed set of experiments that  analyze the effects of domain, forward model, and noise-level shifts in natural and medical image recovery problems using deep unrolled methods. 
\end{itemize}

\section{Related Work}
\subsection{Inverse problems and structured priors.} 
Hand-crafted priors for inverse problems usually assume the signal is sparse in some transform domain. $\ell_1$ norm has been widely used as a sparsity-promoting regularizer \cite{candes2006robust, donoho2006compressed}. Total variation (TV) minimization is used as a regularization approach in \cite{beck2009fast, RUDIN1992259} to solve denoising and deblurring problems. An iterative algorithm that minimizes the image total variation (TV) for CT reconstruction was proposed in \cite{sidky2006accurate}. These hand-crafted priors, however, have limited ability to represent the true underlying image and may lead to sub-optimal solutions \cite{qayyum2022untrained}. 

\subsection{Deep networks for inverse problems}
Generative models learn to map a low-dimensional code into an image. Following \cite{bora2017compressed}, several methods have successfully applied generative networks as priors when solving inverse problems including MRI compressed sensing \cite{NEURIPS2021_7d6044e9}, super-resolution \cite{menon2020pulse}, blind image deconvolution \cite{asim2020blind}, and phase retrieval \cite{hand2018phase, hyder2019alternating}.  

End-to-end trained networks are purely data driven methods that learn to directly map measurements to signals. A denoising network that directly maps corrupted images to clean images was proposed in \cite{zhu2018image}. The method was applied to MRI measurements captured under different acquisition setups \cite{lehtinen2018noise2noise}. Other approaches such as \cite{sriram2020end, jin2017deep} use end-to-end networks to estimate artifact free signals from initial states. 

Plug-and-play (PnP) methods are at the intersection of data driven and model based methods that alternatively minimizes data consistency and regularization terms. PnP-ADMM \cite{venkatakrishnan2013plug} was the first plug-and-play iterative algorithm that used pre-trained denoisers as priors. This method is based on the ADMM algorithm \cite{boyd2011distributed}. PnP-FISTA \cite{kamilov2017plug} is a PnP variant that replaces the proximal operator \cite{boyd2011distributed} of the data fidelity with the gradient. These methods have been applied to solve inverse problems \cite{ chan2016plug, ahmad2020plug}

Deep unrolled networks learn the denoiser network in PnP algorithms in a supervised manner \cite{zhang2018ista, monga2021algorithm, liu2021recovery}. These methods truncate the PnP algorithm for a fixed number of iterations and share the same network through the iterations. They perform updates using the reconstruction output of the final iteration. Deep unrolled methods show remarkable results in several inverse problems such as super-resolution \cite{Zhang_2020_CVPR}, image restoration \cite{Mou_2022_CVPR}, MRI \cite{jun2021joint} and CT \cite{wu2019computationally} reconstruction.

\subsection{Domain expansion and adaptation}
Developing a single network that can handle multiple domains as well as adapt to new target domains has been an active area of research. Deep neural networks can learn transferable features and fine-tuning to a new dataset improves generalization performance \cite{yosinski2014transferable,long2018transferable}. Despite its success, fine-tuning a network or parts of it force the network to lose previously learned domain or task, which requires storing multiple networks per domain and task. Parameter-efficient fine-tuning methods \cite{hu2021lora, He_Li_Zhang_Yang_Wang_2023, NEURIPS2022_efb02f96} propose networks that can achieve competitive performance to fully-tuned networks while requiring few number of additional parameters. Adapter-based techniques that learn efficient modules have been proposed in \cite{Rebuffi_2018_CVPR, li2022cross, chen2022adaptformer}. These modules are added to a pretrained network and enable it to adapt to new tasks. 

Domain specific sub-network selection using binary masks was proposed in \cite{NEURIPS2020_ad1f8bb9, mallya2018piggyback}. 
\cite{shaker2022modular} proposed a modular-network that learns new tasks without compromising performance on previous tasks. The proposed method was successfully applied to a rehearsal-based continual learning method. Such methods, however, require a replay buffer, which is a subset of training samples from previous tasks. A modular-network for continuous task adaptation that does not require replay buffers was proposed in \cite{veniat2021efficient}. Up on arrival of a new task/domain, the method creates trainable modules at every layer and finds the optimal way to add them to a frozen backbone network. These added modules are required to match the base-network in terms of parameters. After training, modules that are not part of the optimal path way will be discarded. This method is computationally demanding and parameter inefficient. 
Later, we will show that modules with significantly fewer parameters compared to the base-network modules we can perform successful task/domain adaptation.

Tuning specific layers such as the BatchNorm \cite{frankle2020training}, the final classification head \cite{he2020momentum}, and LayerNorm \cite{basu2023strong} are proven to be effective adaptation techniques. A related approach that scales and shifts features to achieve the performance of full-tuning was proposed in \cite{lian2022scaling}. In \cite{kanakis2020reparameterizing, rosenfeld2018incremental}, a network reparametrization technique was proposed to learn shared and task-specific modules, enabling a single network to adapt to various settings. Hyperdomain Networks \cite{alanov2022hyperdomainnet} use modulated convolution to adapt generator networks to new domains. 
An adaptation method for shifts in domain and forward-models when solving inverse problems was proposed in \cite{gilton2021model}. The method proposes a fine-tuning and regularization technique adopted from RED \cite{romano2017little}. Domain-specific batch normalization layers were proposed in \cite{karani2018lifelong} for a segmentation network that can handle brain MR scans across scanners and protocols. Unlike R\&R \cite{romano2017little}, the method proposed in \cite{karani2018lifelong} can adapt to new domains without forgetting previous domains. Several test-time adaptation techniques have been proposed to close performance gaps resulting from domain shifts \cite{darestani2022test, Song_2023_WACV, NEURIPS2022_28e9eff8}. While many of these methods are proposed for purely data drive approaches, we focus on methods that fuse data-driven and model based techniques. In addition, our aim is to find parameter efficient domain adaptation techniques without introducing catastrophic forgetting.

\section{Methods}
In this section, we present details of our proposed domain expansion method for deep unrolling-based reconstruction. We first briefly discuss deep unrolled networks (readers may refer to \cite{kamilov2023plug} for further details). Then we discuss how we adapt the network weights using rank-one factors to perform domain expansion/adaptation.

\subsection{Deep unrolled network}

A deep unrolled network in its simplest form represents a fixed number of iterations for solving the optimization problem in \eqref{eq:opti_prob}. Plug and play (PnP) methods based on accelerated proximal gradients \cite{venkatakrishnan2013plug,parikh2014proximal,kamilov2023plug} offer a flexible and efficient framework for solving such problems. Key steps of PnP with a deep denoiser at iteration $k$ can be described as follows. 
\begin{align}
    \label{eq:du_iter_data_cons_update}
    &\mathbf{z}^k = \mathbf{x}^{k-1} - \gamma \nabla g(\mathbf{x}^{k-1})\\
    \label{eq:du_iter_reg_update}
    &\mathbf{s}^k = \mathcal{D}(\mathbf{z}^{k}; \theta) \\ %
    \label{eq:du_iter_state_update}
    &\mathbf{x}^k = \mathbf{s}^k + \beta_k (\mathbf{s}^k - \mathbf{s}^{k-1}),
\end{align}
where $\gamma$ is the step size, superscript $k=1,\dots, K$ denotes iteration number, $\nabla g(\cdot)$ denotes gradient of data fidelity with respect to $\mathbf{x}$, $\mathcal{D}(\cdot; \theta)$ denotes a denoiser or artifact removal network with weights $\theta$, $\beta_k  = {(q_{k-1} - 1)}/{q_k}$, and $q_k = (1/2)(1 + \sqrt{1 + 4q_{k-1}^2})$. We can initialize the estimate as $\mathbf{x}^0 = \mathbf{A}^H\mathbf{y}$, where $\mathbf{A}^H$ denotes Hermitian transpose of the measurement operator. 
Similar to \cite{liu2021recovery}, we implement $\mathcal{D}$ as an artifact removal network: $\mathcal{D}(\mathbf{x}; \theta) = \mathbf{x} - \mathbf{f}(\mathbf{x}; \theta)$, where $\mathbf{f}$ is a DnCNN-based residual network \cite{zhang2017beyond}. 

We can view each iteration of PnP as one layer of the unrolled network with predefined parameters. 
The output of an unrolled network with denoiser $\mathcal{D}(\cdot, \theta)$ and $K$ iterations can be denoted as $\mathbf{x}^K(\theta)$. Since all operations are differentiable, we can further improve the performance by minimizing the reconstruction error on some training images with respect to $\theta$. We can define such an optimization problem as 
\begin{equation}
    \label{eq:du_optim_prob}
\min_{\theta} \sum_{\mathbf{x} \in \mathcal{X}} \mathcal{L} (\mathbf{x}, \mathbf{x}^K(\theta)),
\end{equation}
where $\mathcal{X}$ denotes the set of training images.

\subsection{Factorized network adaptation}
Our method primarily adapts the prior in the unrolled network using domain/task-specific rank-one factors as the data, measurement, or noise distribution changes. We start with a pretrained network $\mathcal{D}(\cdot; \theta)$ with parameters $\theta$. Then we learn domain-specific modulations denoted as $\{M_d\}_{d=1}^{D}$ for $D$ domains. Each $M_d$ represents a set of domain-specific modulation parameters that we use to adapt base network parameters to $\theta \odot M_d$, where $\odot$ represents element-wise multiplication. In order for this multiplication to be defined, we require $\theta$ and $M_d$ to have identical number of elements. In practice, we do not create a new set of modulated weights; instead we keep the $M_d$ and $\theta$ separate. This allows us to fix the base network and adapt to multiple new domains without forgetting previous domains. We represent the domain-specific network for $d$th domain as $\mathcal{D}(\cdot, \theta, M_d)$ and the output of the unrolled network as $\mathbf{x}^K(\theta, M_d)$. To learn the modulation parameters for $d$th domain, we keep $\theta$ unchanged and solve the following optimization problem for $M_i$: 
\begin{equation}
    \label{eq:du_opti_prob_modulated}
\min_{M_d} \sum_{\mathbf{x} \in \mathcal{X}_d} \mathcal{L} (\mathbf{x}, \mathbf{x}^K(\theta, M_d)),
\end{equation}
where $\mathcal{X}_d$ denotes the set of training images for the $d$th domain.

Even though we do not explicitly discuss measurement operator $\mathbf{A}$ and noise $\eta$ in the unrolled network, any mismatch between training and test time settings of domain, measurements, and noise can cause performance degradation. We can consider any variation in data, measurements, or noise as a new domain and use the same procedure described above to learn the domain-specific modulations.

\begin{algorithm}[tb]
\caption{Factorized network adaptation}
\label{alg:algorithm_unrolled_opt}
\textbf{Input}: Training images $\mathbf{x} \in \mathcal{X}_d$ with measurements $\mathbf{y}$, and operator $\mathbf{A}$ for domain indicator $d$ \\ 
Base network parameters $\theta$, $\{\beta_k\}_{k \geq 0}$, $\gamma$, $\alpha$ \\
\textbf{Output}: Recovered image $\mathbf{x}^K$ and domain-specific $M_d$ 
\begin{algorithmic}[1] %
\STATE $M_d \leftarrow \texttt{initialModulation}(d)$
\REPEAT{}
    \STATE for every $\mathbf{x} \in \mathcal{X}_d$ and $ \mathbf{y}$\\ 
    initialize $\mathbf{x}^0 \leftarrow \mathbf{A}^H\mathbf{y}$ 
    \FOR{$k \in \{1,\dots,K\}$}
        \STATE $\mathbf{z}^k \leftarrow \mathbf{x}^{k-1} - \gamma \nabla g(\mathbf{x}^{k-1})$
        \STATE $\mathbf{s}^k \leftarrow \mathcal{D}(\mathbf{z}^{k}; \theta, M_d)$
        \STATE $\mathbf{x}^k \leftarrow \mathbf{s}^k + \beta_k ( \mathbf{s}^k - \mathbf{s}^{k-1} )$
    \ENDFOR
\STATE Calculate loss for all training samples in a minibatch and compute gradient w.r.t. $M_d$\; 

\STATE $M_d \leftarrow M_d - \alpha \nabla_{M_d} \sum_{\mathbf{x} \in \mathcal{X}_d} \mathcal{L}(\mathbf{x}^{K}, \mathbf{x})$\;
\UNTIL{Convergence of $M_d$}
\STATE \textbf{return} $\mathbf{x}^K, M_d$
\end{algorithmic}
\end{algorithm}

\noindent \textbf{Rank-one factorization.}
Inspired by \cite{li2018measuring, hu2021lora}, we assume the intrinsic dimension of the objective in \eqref{eq:du_opti_prob_modulated} is small. We parameterize $M_d$ such that its trainable parameters remains significantly smaller than the number of parameters in the base network.

To achieve the goal of parameter efficiency, we represent modulation weights for each layer as a rank-one tensor. 
Let us assume $l$th convolution layer has weights $W^l$ with kernels of size $k\times k$ with $C_{in}$ input and $C_{out}$ output channels. We represent the modulation weights for $d$th domain and $l$th layer as an outer product of four vectors as 
\begin{equation}
    \small
    \label{eqn:low_rank_combine}
    M^{l}_d = M_d^{1,l} \otimes M_d^{2,l}\otimes M_d^{3,l} \otimes M_d^{4,l},
\end{equation}
where $M_d^{1,l}\in \mathbb{R}^k, M_d^{2,l}\in \mathbb{R}^k, M_d^{3,l} \in \mathbb{R}^{C_{in}}, M_d^{4,l}\in \mathbb{R}^{C_{out}}$. Thus, we need $k+k+C_{in}+C_{out}$ parameters to adapt a layer with $k^2C_{in}C_{out}$ parameters. 
We apply the rank-one factorization and modulation on the convolution layers as follows. 
For an input $U$ with $C_{in}$ channels, we can represent $i$th output channel of the convolution layer as
\begin{equation}
    \small
    \label{eqn:basic_convolution}
    V(:,:,i) = \sum_{j=1}^{C_{in}}  W^l(:,:,j,i) * U(:,:,j),
\end{equation}
where $*$ represents 2D convolution. 
Modulated weights for domain $d$ and layer $l$ can be represented as $W^l_d = W^l \odot M_d^l$. We can represent the convolution operation as 
\begin{equation}
    \small
    \label{eqn:modulated_convolution}
    V(:,:,i) = M_d^{4,l}(i)\left[\sum_{j=1}^{C_{in}} \widetilde W^l(:,:,j,i) * \widetilde U(:,:,j)\right],
\end{equation}
where $\widetilde W^l(:,:,j,i) = W^l(:,:,j,i)\odot (M_d^{1,l}\otimes M_d^{2,l})$ represents a modulated version of $(j,i)$ slice of weight tensor
and 
$\widetilde U(:,:,j) = U(:,:,j) \odot M_d^{3,l}$ represents a modulated version of the $j$th input channel. 
In summary, even though we represent modulation weights are rank-one tensor, we do not need to modulate the weights of the base network. We can implement the same procedure by modulating input channels, 2D filters, and output channels.

\begin{figure}[!t]
    \centering
    \includegraphics[width=\columnwidth]{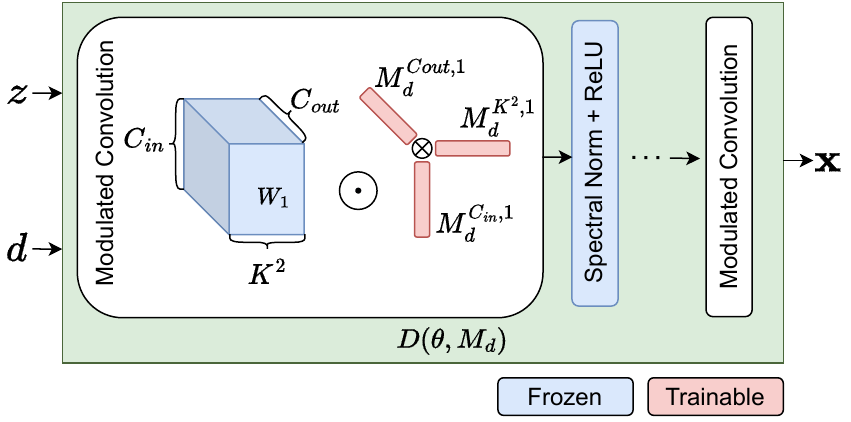}
    \caption{ Overview of our factorized network that uses modulated convolutions for domain adaptation. Our network follows the DNCNN \cite{zhang2017beyond} architecture that leverages modulated convolution for domain adaptation. After trained on the source domain, the network learns low-rank modulations for each domain while keeping the base network parameters frozen. Using a domain identifier, the network selects the appropriate low-rank factors during inference and applies them to the pretrained network through element-wise multiplication. }
    \label{fig:proposed_method}
\end{figure}

Figure \ref{fig:proposed_method} illustrates how our proposed unrolled multi-domain network applies low-rank factors to the pretrained network. We implement \eqref{eqn:modulated_convolution} by first combining the low-rank factors as formulated in \eqref{eqn:low_rank_combine} and applying them to the base convolution weights using an element-wise product. We then use these updated weights to perform regular convolution during the forward pass. When performing backward propagation, we compute gradients with respect to the low-rank factors and update them while keeping the remaining parameters of the network frozen.

A pseudocode for factorized adaptation with the unrolled network is provided in Algorithm~\ref{alg:algorithm_unrolled_opt}. The algorithm begins by initializing domain-specific modulations using an outer product of the low-rank factors. These low-rank factors are real-valued and randomly initialized. After computing the initial estimates $\mathbf{x}^0$, we perform $K$ unrolled iterations containing data-consistency and artificial-removal updates. Finally, we use the output from the last iteration, $\mathbf{x}^K$, to compute the reconstruction loss. This loss is used to compute gradients with respect to the low-rank factors and to perform updates. Further details and hyper-parameter setups are provided in the supplementary material.

\section{Experiments and Results}
We performed a number of experiments to analyse the effects of shifts in different parts of the inverse problem in  \eqref{eq:fwd_model}. The shifts can occur in the data distribution $\mathbf{x}$, the forward model $\mathbf{A}$, and the measurement noise $\eta$. We test our proposed adaptation technique for all these shifts. In all our experiments, we start with a fixed base network, which we refer to as Base AR, and learn domain-specific rank-one modulations. Base AR is trained to reconstruct MR images from $4\times$ radially sub-sampled Fourier measurements without any measurement noise. 
Base AR uses spectral normalization proposed in \cite{miyato2018spectral} along with the ReLU activation functions. We implement our AR network using a 12-layer DnCNN \cite{zhang2017beyond} network. We will provide training details as well as hyper-parameters used in our experiments in the supplementary material.

\begin{figure}[!t]
    \centering
    \includegraphics[width=\columnwidth]{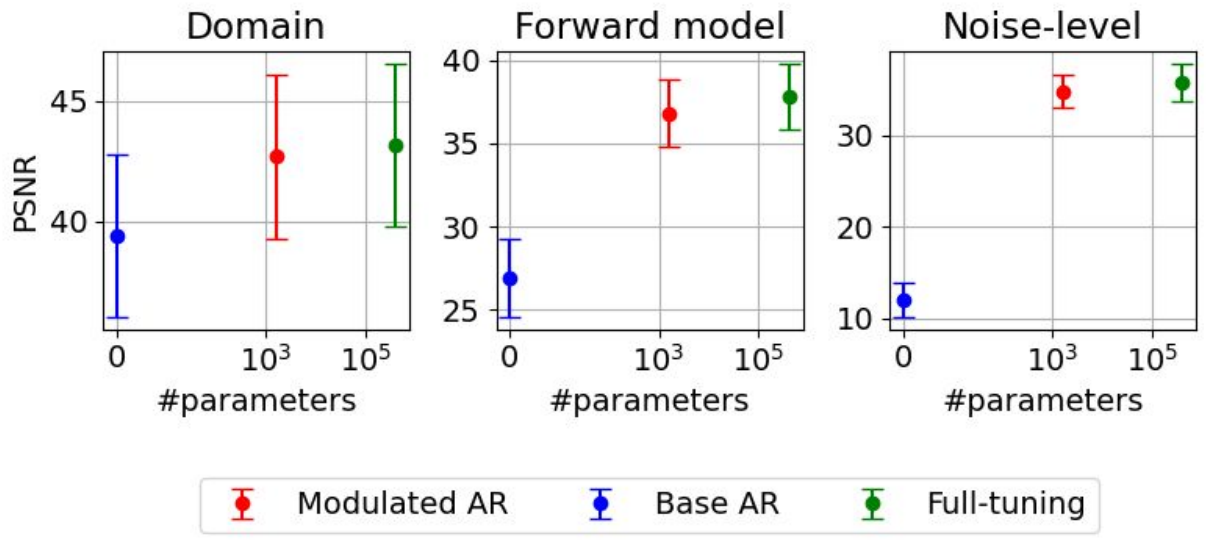}
    \caption{Comparison of our modulated AR, fully-tuned AR, and the Base AR networks in terms of accuracy and number of additional parameters they require. Base AR requires no additional parameter and provides worst performance. Fully-tune AR provides best performance using a large number of parameters. Our proposed method, Modulated AR, shows performance comparable to Fully-tuned AR with a fraction of additional parameters. }\label{fig:num_params_comparison}
\end{figure}
\subsection{Parameter efficiency for adaptation} Figure~\ref{fig:num_params_comparison} compares the performance of a base network, full training, and our proposed modulation-based adaptation for shifts in data distribution/domain, forward model, and noise level. Base network does not require any additional parameter for different domains, but it provides worst performance. Full training learns a new network for every domain/distribution shifts and provides best performance, but at the expense of a large number of parameters per domain. Our proposed network adaptation approach requires a small number of parameters (nearly 1.6K additional parameters) and achieves performance close to full training method. The additional parameters are unique for each domain and are stored separately from the base network. In this manner, the pre-trained model can be adapted to learn new domains while retaining previously learned knowledge.

\begin{figure}[ht]
    \centering
    \includegraphics[width=\columnwidth]
    {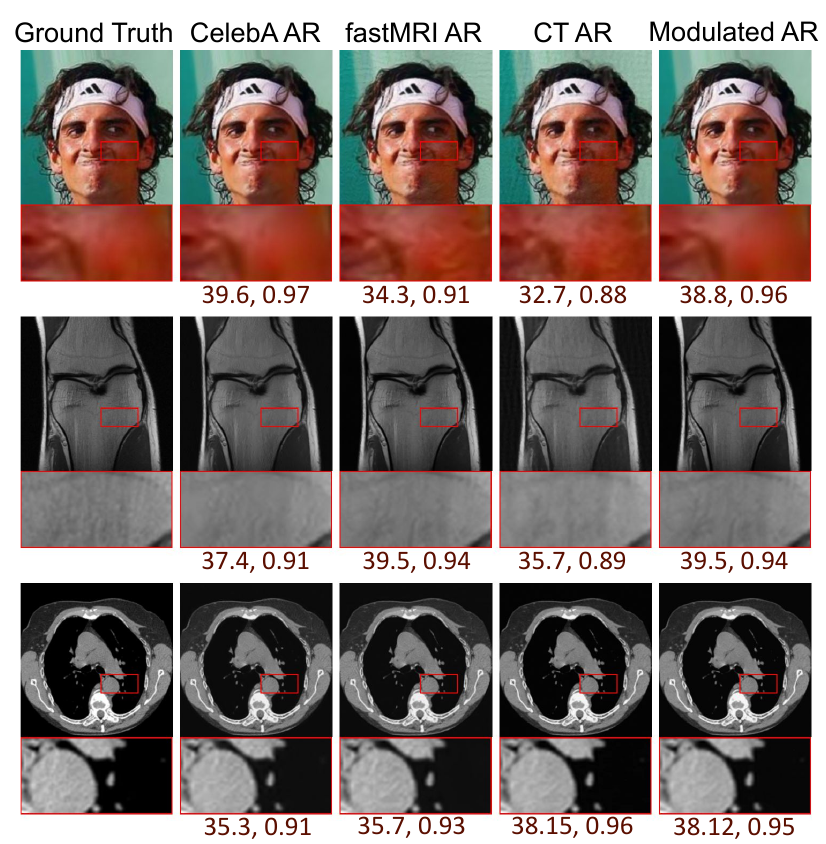}
    \caption{We present sample ground truth images in the first column and reconstruction of these images using three AR networks trained on Face, MR, and CT images in the subsequent three columns .
    Our modulated AR, shown in the last column effectively removes this artifacts and closes the performance gap.}
    \label{fig:domain_shift_all_exps}
\end{figure}
\begin{figure}[ht]
    \centering
    \includegraphics[width=0.9\columnwidth]{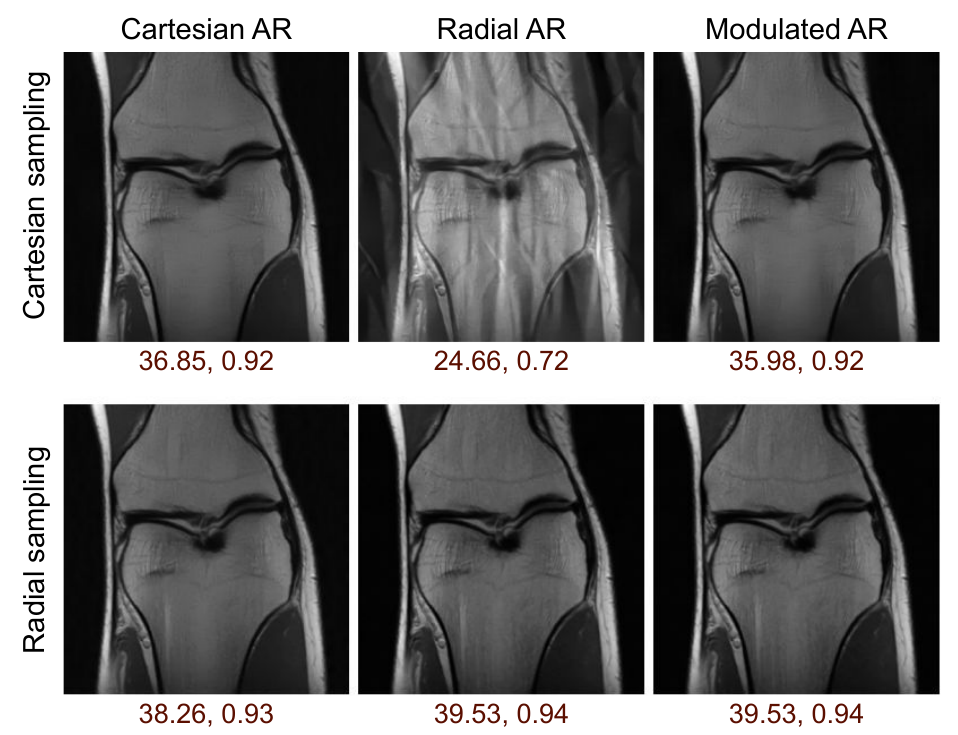}
    \caption{Reconstruction results under sampling pattern shifts. AR trained on radial pattern performs poorly when tested on Cartesian sampled patterns. Our Modulated AR applies low-rank modulations to adapt Radial AR to Cartesian samples. }
    \label{fig:sampling_pattern_shift_exps}
\end{figure}

\begin{table*}[t]
    \small
    \centering
    \caption{Average PSNR of AR networks under domain shift. ARs trained for specific domain (MRI, CelebA, and CT ARs) do not perform well on out-of-domain samples. In contrast, our Modulated AR network, that applies learned modulations for each target domain, has the best average performance across all domains. }
    \label{tab:results_domain_shift}
    \resizebox{0.7\textwidth}{!}{
    \begin{tabular}{ccccc}
    \hline
    Test domain &
      \begin{tabular}[c]{@{}c@{}}AR Trained on\\  MRI\end{tabular} &
      \begin{tabular}[c]{@{}c@{}}AR Trained on\\ CelebA\end{tabular} &
      \begin{tabular}[c]{@{}c@{}}AR Trained on \\ CT\end{tabular} &
      \begin{tabular}[c]{@{}c@{}}Modulated AR\\ (Ours)\end{tabular} \\ \hline
    MRI    & \textbf{40.93} & 39.22 & 37.14&40.93 \\
    CelebA & 40.34 & \textbf{44.29} & 35.44 & 42.97 \\
    CT     & 37.68 & 38.56 & 41.97 &  \textbf{42.25} \\ \hdashline
    Avg    & 39.65 & 40.69 & 38.18 & \underline{42.05} \\ \hline
    \end{tabular}
    }
\end{table*}

\begin{table*}[ht]
\centering
    \caption{Comparison of our method with existing domain adaptation techniques. We have highlighted the best-performing method in boldface and second best with an underscore. Additionally, we provide the count of additional parameters required by each method. Our modulated AR outperforms other methods and is comparable to full tuning. 
    }
    \label{tab:domain_shift_comp_results}
\resizebox{0.7\textwidth}{!}{%
    \begin{tabular}{ccccccc}
    \hline
    Target domain&
      Full-tuning &
      Supsup  &
      RCM   &
      Hyperdomain &
      \begin{tabular}[c]{@{}c@{}}Modulated AR\\ (Ours)\end{tabular} \\
     &
      407k &
      407k &
      50.6k &
      0.7k &
      1.6k \\ \hline
    CelebA  &   \textbf{44.29} & 42.49 & \underline{43.88} & 42.73 & 42.97 \\
    CT      &   \underline{41.97} & 40.57 & 40.99 & 41.57 & \textbf{42.25} \\ \hdashline
    Avg     &   \textbf{43.13} & 41.53 & 42.49 & 42.15 & \underline{42.61} \\ \hline
    \end{tabular}%
    }
\end{table*}

\subsection{Domain shift}
For experiments with domain/data distribution shifts in $\mathbf{x}$, we consider natural image, MRI, and CT scans. We use CelebA dataset \cite{liu2015faceattributes} for natural images, NYU fastMRI dataset for \cite{zbontar2018fastMRI} knee MRI scans, and a subset of TCGA-LUAD dataset \cite{Clark_2013} for CT scans. The first three columns of Table~\ref{tab:results_domain_shift} show the performance single domain AR networks. We present the reconstruction PSNR of these AR networks evaluated under the domain shifts. The last column shows the performance of our modulated network that uses weights of the Base AR trained on MRI and learned modulations for each target domain. Quantitatively we observe that performance drops as domains change (off-diagonal entries in columns 2,3,4). Our proposed method for modulated AR offers best overall performance. Figure~\ref{fig:domain_shift_all_exps} shows example reconstructed images for our domain shift experiments. Our modulated network effectively removes artifacts introduced by fastMRI AR and CT AR on CelebA images.

\noindent \textbf{Comparison with existing domain adaptation methods.} We compare our proposed approach with the following related domain adaptation techniques: Supsup \cite{NEURIPS2020_ad1f8bb9}, RCM \cite{kanakis2020reparameterizing}, Hyperdomain Modulation \cite{alanov2022hyperdomainnet}, and Full-tuning. We evaluate these methods using the same training and testing procedure as our proposed approach. Supsup \cite{NEURIPS2020_ad1f8bb9} learns binary masks to find domain specific sub-networks. RCM \cite{kanakis2020reparameterizing} reparameterizes convolutions using domain-specific feature transformations. Hyperdomain \cite{alanov2022hyperdomainnet} learns domain-specific modulation for input channel of every convolution operation.  Full-tuning retrains the entire network for each target domain and is considered as an upper-bound. Table~\ref{tab:domain_shift_comp_results} shows comparison of these methods and our proposed method outperforms other adaptation techniques while requiring fewer additional trainable parameters.

\begin{table*}[htbp]
\centering
\caption{Sampling pattern shift adaptation results. Our Modulated AR achieves competitive in-domain performance to ARs trained on specific patterns. Moreover, it shows an overall superior performance across all patterns.}
\label{tab:results_sampling_pattern_shift}
\resizebox{0.7\textwidth}{!}{%
\begin{tabular}{cccccc}
\hline
Test pattern &
  \begin{tabular}[c]{@{}c@{}}Radial  \\ AR \end{tabular} &
  \begin{tabular}[c]{@{}c@{}}Cartesian\\AR   \end{tabular} &
  \begin{tabular}[c]{@{}c@{}}Gaussian \\AR  \end{tabular} &
  \begin{tabular}[c]{@{}c@{}}Spiral \\ AR  \end{tabular} &
  \begin{tabular}[c]{@{}c@{}}Modulated AR \\  (Ours)\end{tabular} \\ \hline
Radial    & \textbf{40.93} & 37.75 & 40.55 & 40.83 & \textbf{40.93} \\
Cartesian & 29.74 & \textbf{39.21} & 28.39 & 29.12 & 37.10 \\
Gaussian  & 41.91 & 40.19 & 42.05 & 42.04 & \textbf{42.10} \\
Spiral    & 41.24 & 39.57 & 41.26 & 41.36 & \textbf{41.38} \\ 
\hdashline
Avg       & 38.46 & 39.18 & 38.06 & 38.34 & \textbf{40.38} \\ \hline
\end{tabular}%
}
\end{table*}

\begin{table}[t]
\centering
\small
\caption{Sampling ratio shift adaptation results.}
\label{tab:results_sampling_ratio_shift}
\resizebox{\columnwidth}{!}{%
\begin{tabular}{ccccc}
\hline
Test ratio &
  \begin{tabular}[c]{@{}c@{}}4x AR\end{tabular} &
  \begin{tabular}[c]{@{}c@{}}8x AR\end{tabular} &
  \begin{tabular}[c]{@{}c@{}}10x AR\end{tabular} &
  \begin{tabular}[c]{@{}c@{}}Modualted AR\\ (Ours)\end{tabular} \\ \hline
4x  & \textbf{40.93} & 40.23 & 39.61 & \textbf{40.93} \\
8x  & 34.98 & 37.13 & 37.05 & \textbf{37.32} \\
10x & 31.00 & 33.63 & \textbf{35.34} & 34.73 \\ \hdashline
Avg & 35.64 & 37.00 & 37.33 & \textbf{37.66} \\ \hline
\end{tabular}%
}
\end{table}

\subsection{Forward model shifts}
To evaluate the performance with shifts in the forward model, $\mathbf{A}$, we consider sampling types, ratio, and patterns as domains that can induce shifts. The sampling type can be either Fourier or Gaussian sampling. In the case of Fourier sampling, we can have Cartesian, Radial, Gaussian, or Spiral patterns. The sampling ratio determines the rate at which measurements are captured. We consider reconstruction from $4\times$, $8\times$ and $10\times$ under-sampled measurements. We will now examine the effects of each of these shifts and utilize our proposed method to adapt our Base AR.

\noindent \textbf{Sampling pattern shifts.} Table~\ref{tab:results_sampling_pattern_shift} shows the performance of AR networks trained on single sampling patterns when tested on all available patterns in the first four columns. The last column shows the performance of our modulated AR. We observed a significant performance drop when our Base AR was tested on samples from Cartesian samples. This drop is also evident qualitatively in Figure \ref{fig:sampling_pattern_shift_exps}, where visible artifacts appear in the output. Our modulated AR successfully eliminates these artifacts and bridges the performance gap. Moreover, our method provides overall superior performance compared to networks trained for  individual patterns.

\begin{table*}[htbp]
    \centering
    \caption{Comparison of domain adaptation methods under forward model shifts. Our proposed method achieves competitive performance to full-tuning with significantly fewer parameters. It outperforms related domain adaptation methods in terms of performance and parameter efficiency.}
    \label{tab:forward_model_shift_comp_results}
    \resizebox{0.7\textwidth}{!}{%
    \large
    \begin{tabular}{ccccccc}
    \hline
    \begin{tabular}[c]{@{}c@{}}Sampling shifts\end{tabular} & 
      Full-tuning &
      Supsup  &
      RCM   &
      \begin{tabular}[c]{@{}c@{}}Hyperdomain \end{tabular} &
      \begin{tabular}[c]{@{}c@{}}Modulated AR\\ (Ours)\end{tabular} \\ &
      407k &
      407k &
      50.6k &
      0.7k &
      1.6k \\ \hline
    \begin{tabular}[c]{@{}c@{}} Radial to  Cartesian\end{tabular} &
      \textbf{39.21} &
      36.37 &
      36.52 &
      36.57 &
      \underline{37.10} \\
    \begin{tabular}[c]{@{}c@{}} 4x to 10x\end{tabular} &
      \textbf{35.34} &
      33.62 &
      34.49 &
      34.56 &
      \underline{34.73} \\
    \begin{tabular}[c]{@{}c@{}} Fourier to  Gaussian\end{tabular} &
      \textbf{38.59} &
      36.31 &
      38.49 &
      38.45 &
      \underline{38.55} \\ \hdashline
    Avg &
      \textbf{37.71} &
      35.43 &
      36.50 &
      36.53 &
      \underline{36.79} \\ \hline
    \end{tabular}%
    }
\end{table*}

\begin{figure}[ht]
    \centering
    \includegraphics[width=\columnwidth]{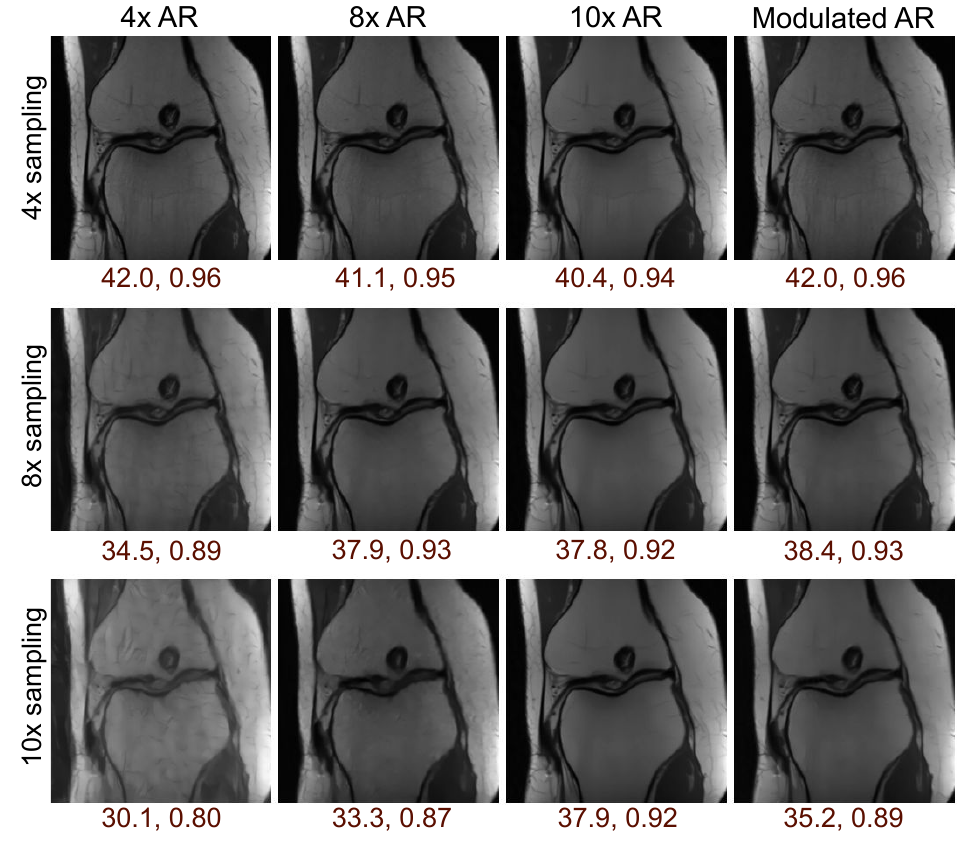}
    \caption{Examples of image reconstruction under sampling ratio shifts. Our Modulated AR shows an average superior performance when compared to the 4×, 8×, and 10× AR networks.
    }
    \label{fig:sampling_ratio_shift_exps}
\end{figure}

\noindent \textbf{Sampling ratio shifts.}  We compared the performance of different AR networks trained on three sampling ratios and presented the results in Table \ref{tab:results_sampling_ratio_shift}. The $4\times$ AR network exhibits poor performance when tested with $8\times$ and $10\times$ radially subsampled measurements. Additionally, the AR network trained on the $8\times$ ratio did not perform well with $10\times$ ratio. To address this, we applied our modulation technique to adapt the Base AR model to $8\times$ and $10\times$ sampling ratios. On average, the modulated network outperforms AR networks trained on specific sampling ratios. Figure \ref{fig:sampling_ratio_shift_exps} illustrates the reconstruction results of the networks trained at various sampling ratios, including our modulated network. On average, the modulated network outperforms AR networks trained on specific sampling ratios. 

\noindent \textbf{Comparison with existing domain adaptation methods.} We now compare our method with some of the existing domain adaptation techniques under the forward model shifts discussed above. We report the average PSNR along with the number of trainable parameters with in each method in Table~\ref{tab:forward_model_shift_comp_results}. Our proposed method outperforms all domain techniques and is only one dB less than full-tuning, which requires significantly larger number of parameters.

\begin{figure}[thbp]
    \centering
    \includegraphics[width=\columnwidth]{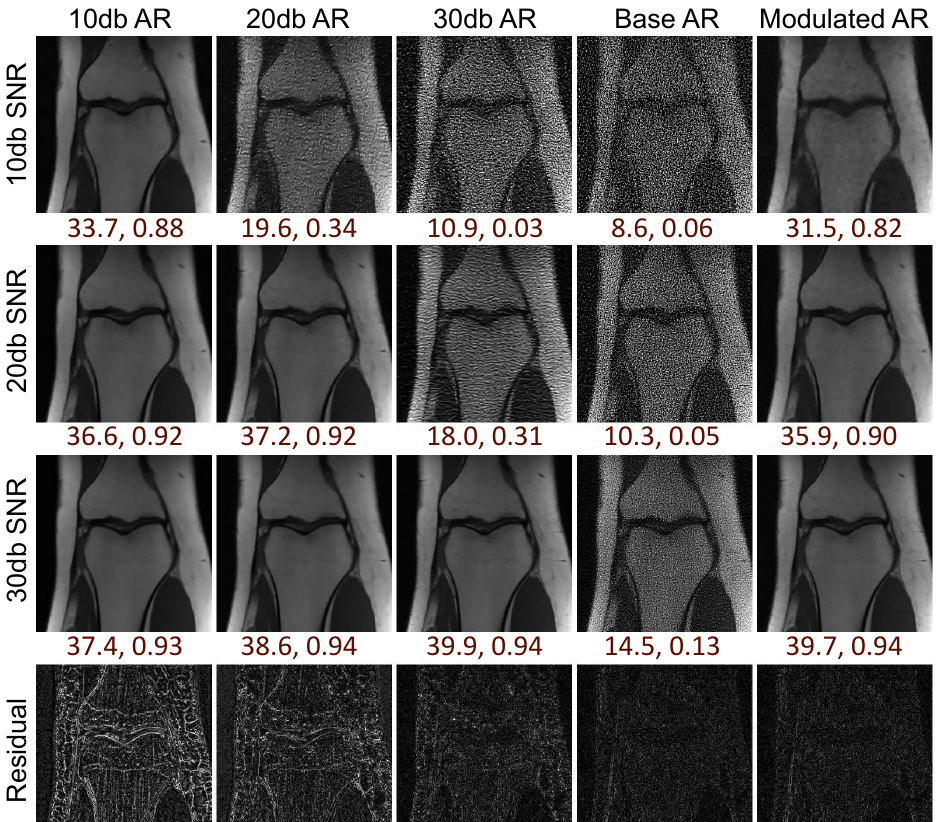}
    \caption{Visual results of models trained at specific noise levels and our modulated network under measurement level. The last row shows the $20\times$amplified residual of the reconstructed image under no measurement noise.}
    \label{fig:snr_shift_exps}
\end{figure}

\begin{table*}[thbp]
\centering

    \caption{Noise level shift adaptation results. 
    Modulated AR that learns low-rank factors for each noise-level outperforms networks trained for a specifc noise.
    }
    \label{tab:noise_level_adaptation_results}
    \resizebox{0.7\textwidth}{!}{%
    \begin{tabular}{cccccc}
    \hline
    \begin{tabular}[c]{@{}c@{}}Test SNR\\ \end{tabular} &
       10db AR &
       20db AR &
       30db AR &
       Base AR &
      \begin{tabular}[c]{@{}c@{}}Modualted AR\\ (Ours)\end{tabular}  \\ \hline
    10db     & \textbf{33.37} & 23.86 & 12.56 & 9.18 & 31.40 \\
    20db     & 35.39 & \textbf{35.82} & 22.86 & 11.20  & 35.09 \\
    30db     & 35.97 & 36.84 & \textbf{38.04} & 15.90 & 37.91 \\
    No noise & 36.10  & 40.93& 40.93&40.93&40.93 \\ \hdashline
    Avg      &   35.21    &   33.30    &   28.11    &   19.30  & 36.33     \\ \hline
    \end{tabular}%
    }
\end{table*}

\subsection{Noise-level shifts}

Noise-level shifts can also cause significant performance degradation in AR networks. We model the noise as an additive Gaussian noise $\eta \sim \mathcal{N}(0, \sigma^2)$ and analyze the effects of different noise levels on the performance. Figure~\ref{fig:snr_shift_exps} shows sample reconstructed images with the Base AR, our Modulated AR, and AR networks trained for 10, 20, and 30 dB SNR. We observed that the Base AR is unable to reconstruct the MRI scans from the noisy measurements. This is also shown quantitatively in Table \ref{tab:noise_level_adaptation_results}, where the performance of the Base AR is severely degraded in the presence of noise. The AR network trained on 10dB SNR performs well on higher noise settings but fails to recover fine details when tested with noise-free or low noise measurements. The last row of Figure \ref{fig:snr_shift_exps} shows the $20\times$ amplified reconstruction residual of each model when reconstructing noise-free measurements. From this row, we can infer that AR networks trained on higher noise-levels fail to recover fine details when tested with lower noise-levels. To the contrary, our Modulated AR has the ability to reconstruct fine details when the measurement noise is low and maintains comparable performance to noise-specific AR as the noise level increases. 

\noindent \textbf{Comparison with existing domain adaptation methods.} Table \ref{tab:noise_level_shift_comp_results} reports comparison of our proposed method with related domain adaptation techniques. Although Full-tuning and Supsup \cite{NEURIPS2020_ad1f8bb9} show slight performance improvement (less than 1 dB), they require a significant number of trainable parameters. Furthermore, Full-tuning does not have the ability to retrain previously learned knowledge. Our method achieves competitive performance to RCM \cite{kanakis2020reparameterizing} while requiring a fraction of the additional trainable parameters.

\begin{table*}[htbp]
\centering
\caption{Comparison of various domain adaptation methods under noise level shifts. Our proposed method can achieve competitive performance to RCM and Full-tuning while requiring significantly fewer number of additional parameters.}
\label{tab:noise_level_shift_comp_results}
\resizebox{0.7\textwidth}{!}{%
\begin{tabular}{ccccccc}
\hline
    Test SNR &
  Full-tuning &
  Supsup &
  RCM &
  \begin{tabular}[c]{@{}c@{}}Hyperdomain \\ modulation\end{tabular} &
  \begin{tabular}[c]{@{}c@{}}Modulated AR\\ (Ours)\end{tabular} \\
     & 407k  & 407k  & 50.6k & 0.7k  & 1.6k  \\ \hline
10db &                                                                    \textbf{33.37} & 31.94 & \underline{32.80} & 29.11 & 31.40 \\
20db &                                                                    \textbf{35.82} & 35.53 & \underline{35.70} & 34.32 & 35.09 \\
30db &                                                                    \underline{38.04} & 37.75 & \textbf{38.08} & 37.58 & 37.91 \\ \hdashline
Avg  &                                                                    \textbf{35.74} & 35.07 & \underline{35.53} & 33.67 & 34.80 \\ \hline
\end{tabular}%
}
\end{table*}

\section{Limitations} 
While our proposed method is able to continuously adapt to new domains, it requires domain selectors/identifiers during inference to apply the correct modulations. In some cases, this is not a major limitation since we can partially infer the domain from the available measurements or context. In principle, we can parameterize the network modulations as a function of the input and construct a multi-domain network that can infer the domain without the need for explicit identifiers. 
Another limitation of our current method and experiments is the incremental adaptation to target domains. We start from a fixed base network and subsequently adapt it to multiple domains independently. We can further improve the efficiency of our method by adapting the network to multiple domains jointly. Achieving rapid and generalized multi-domain adaptation is feasible following meta-learning techniques as outlined in \cite{finn2017model}. We believe that these limitations will serve as inspiration for several future studies.

\section{Conclusion}
We proposed a simple and parameter-efficient method to adapt networks for domain adaptation and expansion. Our method uses a fixed base network and learns separate (domain-specific) rank-one modulation parameters. This capability allows our method to continually learn new domains while retaining previously acquired knowledge. We focused on shifts that arise in solving inverse problems for imaging, including shifts in data distribution, forward model, and noise level. We demonstrated the effectiveness of our approach in adapting to all these shifts.

\bibliographystyle{IEEEtran}
\bibliography{tci_bib}

\clearpage
\section*{\sc Supplementary Material}
\input{appendix}

\end{document}

%% file: appendix.tex
We present additional material and details to complement our main paper. We provide a detailed description of our training and hyper-parameter tuning procedures. Additionally, we present further experiments, analyze the effects of modulation in each layer, and showcase visual results. Finally, we address the limitations of our work and suggest potential directions for future research.

\section{Training details}

We used PyTorch \cite{paszke2019pytorch} to implement our proposed method on a single NVIDIA GeForce RTX 2080 Ti GPU with 12GB memory. Our artifact removal (AR) prior is implemented using a DnCNN \cite{zhang2017beyond} network with 12 blocks. Each block comprises a convolution layer, a spectral normalization layer \cite{miyato2018spectral}, and a ReLU activation layer \cite{agarap2018deep}. Within the intermediate blocks, our convolution consists of $64$ filter kernels with a size of $3\times3$. The number of filters at the input and output layers is set to match the number of features in the target dataset.

The AR operator is implemented as the residual of our DnCNN network, utilizing an $\alpha$-averaged operator similar to \cite{liu2021recovery}. We set $\alpha=0.2$ for all experiments. With this operator, we obtained the best performance when setting the acceleration parameter $q_k=1$ for all $k \geq 1$. We used the ADAM optimizer \cite{kingma2014adam}, setting the learning rate to $10^{-4}$ for the base network weights during full-tuning and $10^{-2}$ for the modulation factors. Our experiments were trained for $100$ epochs. After $50$ epochs, we adjusted the learning rate for the modulations, reducing it by a factor of $2$. We maintained default settings for the remaining optimizer configurations. We reported the results using the best-performing model on the validation dataset. Throughout all experiments, we incorporated $K=33$ unrolled iterations. For MR image reconstruction tasks, we employed a step size of $\gamma=1.5$, while $\gamma=1.2$ was utilized for face image reconstruction tasks. In the comparison experiments, we adapted the source code available on the GitHub pages of RCM \cite{kanakis2020reparameterizing}, Supsup \cite{NEURIPS2020_ad1f8bb9}, and Hyperdomain Modulation \cite{alanov2022hyperdomainnet} to align with our framework.

We utilized PyTorch's Torchvision library as the data source for CelebA \cite{liu2015faceattributes} and adhered to the official train-val-test set split. For the multi-coil Knee dataset provided by NYU FastMRI \cite{zbontar2018fastMRI}, we partitioned the original training set into training and validation sets using an 85-15 split. Subsequently, we used the validation dataset as our test set. In the case of CT scans, we used a subset of the TCGA-LUAD dataset \cite{Clark_2013}. We performed a 75-15-10 split for the training, validation, and test sets. All images were normalized within the range of $[0,1]$. 

\subsection{Initialization. } To ensure stability, we implement our modulation as $W^l \odot (1 + M_d^l)$. The factors of $M_d^l$ are initialized using a uniform distribution near zero, such that $M_d^{i, l} \sim \textit{U}\left[-\frac{1}{\sqrt{f}}, \frac{1}{\sqrt{f}}\right]$, where $f$ corresponds to the number of channels or kernel size to be modulated at the $l^{th}$ layer. This approach ensures that our modulated weights start with values close to the pre-trained weights and allows them to be updated to an optimal value.

\section{Additional experiments}

\subsection{Analyzing the layer-wise effects of weight modulation}

We conducted an analysis of the effects of weight modulation at each layer. This analysis provides us with better insight into which layers play more crucial roles in bridging performance gaps. Subsequently, we can use this information to apply dynamic adaptation to different layers, thereby further reducing additional computation and the necessary parameters. 

\begin{figure}[htbp]
    \centering
    \includegraphics[width=\columnwidth]{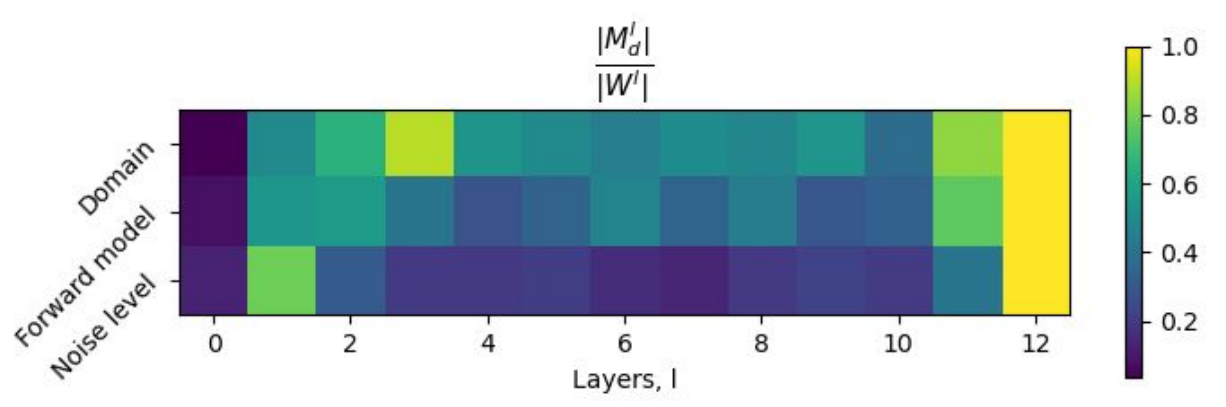}
    \caption{We show most of the modulation occurs at the final layers of our artifact removal network. We present the ratio of $\ell_2$ norms of modulations to weights at each layer for domain, forward model, and noise level shifts. We have normalized each row to enhance visualization.}
    \label{fig:modulation_norms}
\end{figure}

Figure \ref{fig:modulation_norms} illustrates the ratio of $\ell_2$ norms of modulations $|M_d^l|$ to pre-trained weight norms $|W^l|$ at the $l^{th}$ layer, corresponding to domain, forward model, and noise-level adaptation experiments. We normalize the norm ratios independently for each adaptation, scaling them between $0$ and $1$. The figure suggests that modulation power is primarily concentrated in the final layers of the network, with minimal impact observed in the initial layers. 
We confirm this observation by conducting partial modulation experiments for noise-level adaptation, where we apply modulation to a subset of layers in our network. We report our findings in Table \ref{tab:partial_modulation_res}. We note that modulating the top half of the layers (requiring only $0.88$k parameters) resulted in a $7\%$ performance decrease, and modulating the last $4$ layers (requiring only $0.67$k parameters) led to only a $12.8\%$ drop. In contrast, modulating the first half and the first four layers resulted in a significant performance reduction. In future works, we can utilize these insights when designing parameter-efficient domain adaptation techniques.

\begin{table}[htbp]
    \small
    \centering
    \caption{We experimentally demonstrate that modulating the top half and the last four layers of our network retains the majority of the performance achieved through full modulation. In contrast, we show that the same level of performance cannot be attained by solely modulating the initial layers.}
    \label{tab:partial_modulation_res}
    \resizebox{\columnwidth}{!}{%
    \begin{tabular}{lcccc}
    \hline
    Trained blocks & \multicolumn{2}{c}{\begin{tabular}[c]{@{}c@{}}Noise-level \\ Adaptation\end{tabular}} & \multicolumn{2}{c}{\# Trained} \\ %
                & PSNR  & SSIM & Params & Layers \\ \hline
    None         & 9.18  & 0.02 & 1.6k   & 13     \\ 
    All         & 31.40  & 0.79 & 1.6k   & 13     \\ \hdashline
    Blocks 0-4  & 10.67 & 0.04 & 0.67k  & 5      \\
    Blocks 4-8  & 16.87 & 0.20 & 0.67k  & 5      \\
    Blocks 8-12 & 27.15 & 0.49 & 0.67k  & 5      \\ \hdashline
    Blocks 0-6  & 19.97 & 0.28 & 0.88k  & 7      \\
    Blocks 6-12 & 29.14 & 0.64 & 0.88k  & 7      \\ \hline
    \end{tabular}%
    }
\end{table}

\subsection{Adapting to domain and forward model shifts}

\begin{figure}[t]
    \centering
    \includegraphics[width=\columnwidth]{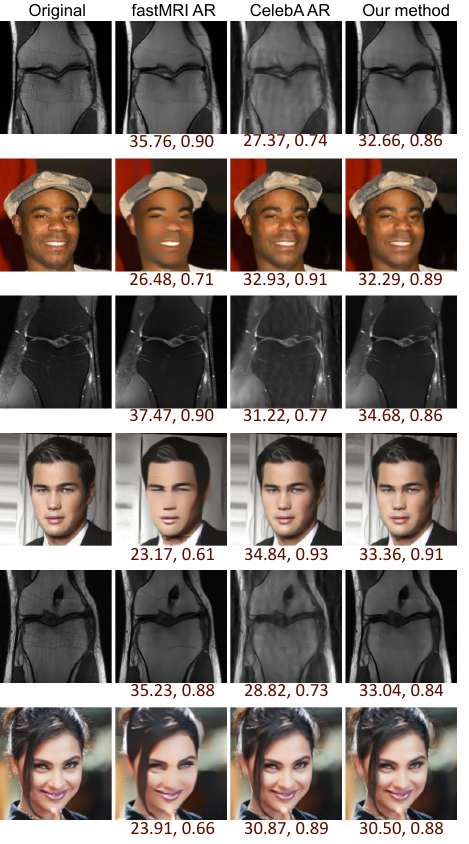}
    \caption{Examples of image reconstruction under domain and forward model shifts. Second and third columns show reconstructed images with fastMRI AR and celebA AR, respectively. Reconstruction quality degrades with domain shifts (PSNR and SSIM reported under each image). Our proposed network adaptation method, where we adapt the mis-matched ARs (recovering MR images using CelebA AR and recovering faces using fastMRI AR) to recover an targets images in shown below.  }
    \label{fig:fwd_and_domain_shift}
    \vspace{10mm}
\end{figure}

In the main paper, we focused on experiments involving a single type of shift that can arise when solving inverse problems. Now, we are expanding this concept by applying our technique to accommodate multiple shifts. Specifically, we will adapt a network trained for face image reconstruction using Gaussian sampling to MR image reconstruction with Fourier sampling, encompassing both domain and forward model shifts. Table \ref{tab:fwd_and_domain_shift_res} shows the performance of these AR networks. We observe an $18\%$ drop in performance on CelebA and a $30.37\%$ drop on FastMRI ARs due to this shift. To address this challenge, we applied our domain adaptation technique and conducted a comparison with relevant methods. We present the average performance in Table \ref{tab:domain_adapt_comp_results}. Similar to our previous experiments, our method outperforms other domain adaptation techniques. Additional visual results for this experiment are shown in figure \ref{fig:fwd_and_domain_shift}.

\begin{table}[h]
    \centering
    \small
    \caption{The CelebA AR, which is trained to reconstruct faces from random projections, exhibits poor performance in MR reconstruction. Similarly, the fastMRI AR is unable to reconstruct face images, despite its competence in its own domain. We provide the average PSNR values for these ARs.}
    \label{tab:fwd_and_domain_shift_res}   
    \begin{tabular}{ccc}
    \hline
    Test dataset & CelebA AR & FastMRI AR \\ \hline
    CelebA & 35.05 & 24.58 \\
    FastMRI & 28.46 & 35.34 \\ \hline
    \end{tabular}
\end{table}

\begin{table}[h]
    \centering
    \small
    \caption{Comparison results upon adapting CelebA AR to MR and CT reconstruction tasks. We have highlighted the best-performing method in boldface and second best with an underscore. }
    \label{tab:domain_adapt_comp_results}  
    \begin{tabular}{lrcc}
    \hline
    Methods & \multicolumn{1}{c}{\begin{tabular}[c]{@{}c@{}}\# Trainable \\ parameters\end{tabular}} & MRI Knee & CT \\ \hline
    Full-tuning & 407k & \textbf{35.34} & \textbf{32.63} \\
    Supsup & 407k & 32.57 & 28.82 \\
    Reparametrized AR & 50.6k & 34.14 & 30.43 \\
    Hyperdomain modulation & 0.7k & 33.81 & 31.37 \\ \hdashline
    Modulated AR & 1.6k & \underline{34.17} &  \underline{32.11}  \\ \hline
    \end{tabular}
\end{table}

\subsection{Adapting to sampling type shifts} 
\begin{figure}[!t]
    \centering
    \includegraphics[width=\columnwidth]{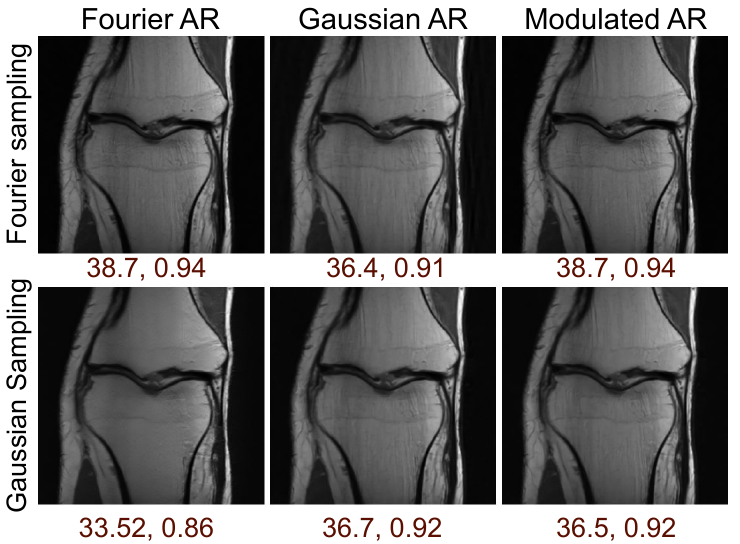}
    \caption{The Fourier (Base) AR is unable to recover fine details under Gaussian sampling (first column, second row). On the other hand, the Modulated AR can achieve performance comparable to fully-tuned networks when reconstructing images from both Fourier and Gaussian samples (last column).}
    \label{fig:sampling_type_shift_exps}
    \vspace{5mm}
\end{figure}

We consider Fourier and Gaussian measurement operators as domains of different sampling types. When testing our Base AR with Gaussian sub-sampled MRI measurements, we observed an average PSNR drop of approximately $2.7$ dBs on our test set. As shown in the second row of the first column of Figure \ref{fig:sampling_type_shift_exps}, the Base AR is incapable of recovering fine details and produces a smoothed output. Using our proposed modulation technique, we were able to enhance the reconstruction results and achieve outcomes similar to those of the fully-tuned AR.

\begin{figure}[!t]
    \centering
    \includegraphics[width=\columnwidth]{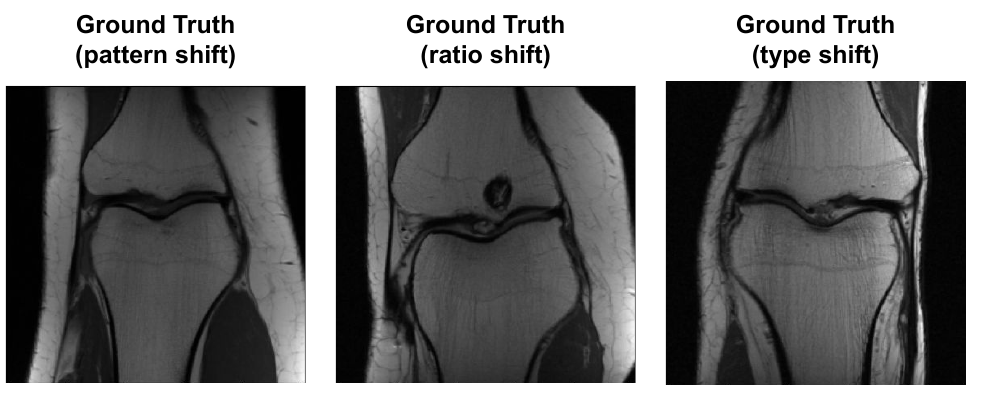}
    \caption{Ground truth images for forward model shifts experiments. }
    \label{fig:ground_truth_images}
    \vspace{80mm}
\end{figure}

Lastly, in Figure \ref{fig:ground_truth_images}, we present the ground truth images for the samples utilized in our forward model adaptation experiments. The corresponding reconstruction outputs of these images are displayed in the main paper.